\documentclass[prd,twocolumn,10pt,aps,longbibliography]{revtex4-2}

\usepackage{graphicx}
\usepackage{dcolumn}
\usepackage{bm}
\usepackage[T2A]{fontenc}
\usepackage{comment}
\usepackage{braket}
\usepackage{amsmath}
\usepackage{amssymb}
\usepackage{dsfont}
\usepackage{upgreek}
\usepackage{natbib}
\newcommand{\bfp}{{\bf p}}
\newcommand{\bfx}{{\bf x}}
\newcommand{\bfq}{{\bf q}}

\DeclareMathOperator{\Tr}{Tr}

\begin{document}
%
\title{Neutrino quantum decoherence in a fluctuating ALPs field}
\author{\firstname{Alexey} \surname{Lichkunov}}
\email[]{lichkunov.aa15@physics.msu.ru}
\affiliation{Department of Theoretical Physics, Faculty of Physics, Lomonosov Moscow State University, 119992 Moscow, Russia, \\ Pushkov Institute of Terrestrial Magnetism, Ionosphere and Radiowave Propagation (IZMIRAN), 108840 Moscow, Troitsk, Russia}
\author{\firstname{Konstantin} \surname{Stankevich}}
\email[]{kl.stankevich@physics.msu.ru}
\affiliation{Department of Theoretical Physics, Faculty of Physics, Lomonosov Moscow State University, 119992 Moscow, Russia}
\author{\firstname{Alexander} \surname{Studenikin}}
\email[]{studenik@srd.sinp.msu.ru}
\affiliation{Department of Theoretical Physics and Department of Physics of Particles and Extreme States of Matter, Faculty of Physics, Lomonosov Moscow State University, 119992 Moscow, Russia}
\begin{abstract}
The evolution of neutrinos in a fluctuating classical axionlike particles (ALPs) field is investigated. The equation for the neutrino density matrix is obtained in the Redfield form, and the corresponding dissipative matrix is derived. Estimates on the ALP-neutrino coupling constants are obtained using the limits on the decoherence parameter determined from the analysis of reactor neutrino experiments data.  
\end{abstract}
\maketitle
\section{Introduction}

The study of neutrino properties provides a pathway to understanding physical processes beyond the Standard Model \cite{RevModPhys.87.531,Giunti:2024gec}. Neutrinos can also be used as a tool for the study of astrophysical objects in the context of multimessenger astronomy \cite{Volpe:2023met,Tamborra:2024fcd}. Additionally, neutrinos can interact with axionlike particles (ALPs) (see, for instance \cite{Cl_ALPs_Neu}), which are one of the most promising candidates for dark matter. This fact can help in the study of dark matter by examining neutrino fluxes from different sources.

axionlike particles are a class of pseudo-Goldstone bosons that arise from spontaneous breaking of the U(1) symmetry \cite{Carenza:2024ehj}. Axions are introduced themselves to solve the strong CP problem. However, unlike axions, the coupling constants and masses of ALPs are independent of axion-pion interactions. Instead, these parameters are free to vary, allowing ALPs to potentially solve various issues in modern particle physics, such as the  discrepancy between the measured value of the muon anomalous magnetic moment, leptonic mass hierarchies, and neutrino mass generation. 

One of the open questions is how to describe the ALPs field. In the literature there is a description of any ALPs as a classical field, since it is assumed that thermalization has occurred in the Universe and any ALP, acquiring mass, is already close to the vacuum of potential energy. In that case ALPs form a Bose-Einstein condensate which oscillates coherently. This approach is described in \cite{Cl_Axion} (for axions) and used in \cite{Cl_ALPs_Neu} for studying neutrino oscillations in a classical ALPs field.

In the present study, we investigate the evolution of Dirac neutrinos in a classical fluctuating ALPs field, which is one of the possible candidates for dark matter. We demonstrate that neutrino interaction with a fluctuating ALP field causes neutrino quantum decoherence. Previously, it was shown that the neutrino quantum decoherence arises from neutrino interaction with fluctuating magnetic field and media \cite{Balantekin,Burgess:1996mz,Benatti:2004hn}, and with stochastic gravitational waves
\cite{Dvornikov:2021sac}. In \cite{MyNew} it was shown that the neutrino quantum decoherence can occur due to the neutrino decay into a lighter neutrino state and a massless particle, as well as due to the inverse process of absorption of a massless particle by a neutrino.

This paper is structured as follows. In Sec. \ref{sec:form}, the formalism of the neutrino quantum decoherence in the presence of fluctuating classical ALPs field is developed. Also here the dissipative matrix for this scenario is obtained. In Sec. \ref{sec:apply} possible conditions of applicability are discussed. In Sec. \ref{sec:est}, estimations on neutrino-ALPs interaction are calculated. For this aim, the limits on the decoherence parameter from \cite{Giunti:2023} are used. Finally, there is a conclusion in Sec. \ref{sec:conc}.
\section{The developed formalism}\label{sec:form}

In order to describe the neutrino evolution, we use the approaches developed in \cite{Balantekin} and \cite{MyNew}. Within this approach, the evolution of the Dirac neutrino is determined by a density matrix

\begin{equation}
    \rho_\nu (t) = \ket{\Phi(t)} \bra{\Phi(t)}
    ,
\end{equation}
where the state $\ket{\Phi(t)}$ describes neutrino with different momenta
\begin{equation}
\ket{\Phi} = \otimes_\bfp\ket{\Phi_\bfp}.
\end{equation}
The stationary neutrino state $i$ with momentum $\bfp$ is given by

\begin{equation}
\ket{i \bfp} = \ket{i} \otimes \ket{\bfp} = \sqrt{2 E_{\bfp i}} a_{\bfp i}^{\dagger} \ket{0}
,
\end{equation}
where $a_{\bfp i}^{\dagger}$ is the creation operator of the neutrino mass state $i$ with momentum $\bfp$ and energy $E_{\bfp i}$. The matrix element of neutrino matrix density is given by 
\begin{equation}\label{diag_p}
    \bra{i \bfp} \rho_\nu(t) \ket{j \bfq} =  (2 \pi)^3 \delta^{(3)}(\bfp-\bfq) \rho_\bfp^{ij}(t)
    .
\end{equation}
Time evolution of density matrix is described by the evolution operator
\begin{equation}\label{dens}
	\rho_\nu \left( t\right) = U(t_0, t) \: \rho_\nu \: U^{\dagger}(t_0, t),
\end{equation}
which is expressed as
\begin{equation}
    U(t_0, t) = Texp \left( -i \int^t_{t_0} H(t') dt' \right)
    ,
\end{equation}
where the Hamiltonian $H(t)$ in the interaction representation is

\begin{equation} \label{Hamiltonian}
	H(t) = \int d^3 \bfx j(x) a(x)
	.
\end{equation}
Here $a(x)$ describes an ALPs field, $j(x) = i\bar {\nu}(x) \left( g_V + g_A \gamma_5 \right) \nu (x)$ is the neutrino current, $g_V$ and $g_A$ are vector and axial coupling constant matrices, which elements are expressed as

\begin{equation}\label{gva0}
    g_V^{ij} = C_{V}^{ij}\frac{m_i - m_j}{F},
\end{equation}
\begin{equation}\label{gva}
    g_A^{ij} = C_{A}^{ij}\frac{m_i + m_j}{F},
\end{equation}
where $m_i$ is a neutrino mass, $F$ is the so-called decay constant and $C_{V}^{ij}$ are dimensionless constants (see, for instance \cite{PhysRevD.99.096005}). A neutrino state has the following form

\begin{equation}
    \nu (x) =  \sum_i \int \dfrac{d^3 \bfp }{(2 \pi)^3} \dfrac{1}{\sqrt{2 E_{\bfp i}}} \hat{a}_{{\bf p} i} u_i(\bfp) e^{-i p x}
    ,
\end{equation}
where the negative-frequency solutions with the antineutrino creation operator $b_{{{\bf p} i}}^\dagger$ were omitted, since we do not consider the neutrino-antineutrino transitions.   

A fluctuating ALPs field $a(x)$ has the following form \cite{PhysRevD.97.123006, Wang:2025}

\begin{equation}
    a(x) = \alpha(t)\frac{\sqrt{2\rho}}{m_a}\cos{(m_at - \bfp\cdot\bfx)},
\end{equation}
where $\rho$ is an ALPs energy density, $m_a$ is an ALPs mass, while $\alpha(t)$ is a non-negative random variable obeying the Rayleigh distribution

\begin{equation}
    f(\alpha) = \alpha \exp{\left(-\frac{\alpha^2}{2}\right)}.
\end{equation}

In the following we consider the particular case of ALPs in Milky Way, where the ALPs velocity is of order $10^{-3}$c \cite{Cl_ALPs_Neu}. Therefore, the dependence on momentum and coordinate can be omitted and
\begin{equation}
    a(t) \approx \alpha(t)\frac{\sqrt{2\rho}}{m_a}\cos{(m_at)}.
\end{equation}

It is possible to decompose ALPs field on the two terms

\begin{equation}
    a(t) = a_{m}(t) + a_{fl}(t).
\end{equation}
The first term $a_{m}(t)$ describes the mean classical ALPs field that oscillates coherently due to the misalignment mechanism

\begin{equation}
    a_{m}(t) = \frac{\sqrt{\pi\rho}}{m_a}\cos{(m_at)},
\end{equation}
while the term $a_{fl}(t)$ describes fluctuations of the classical field

\begin{equation}
    a_{fl}(t) = \left(\alpha(t) -\sqrt\frac{\pi}{2}\right)\frac{\sqrt{2\rho}}{m_a}\cos{(m_at)}.
\end{equation}

To average fluctuating part, we use path integral with Rayleigh measure (similar to \cite{Balantekin}, where Gaussian perturbations are considered). For a function $f(a_{fl})$ the averaging has the following form

\begin{equation}\label{aver}
    \braket{f(a_{fl})} = \int_{0}^{+\infty}\prod_t\left[\alpha(t)d\alpha(t)\right]f(a_{fl})e^{-\int_0^t\frac{\alpha^2(t')}{4\tau}dt'}.
\end{equation}
where $\tau$ is a correlation time.

Therefore, the mean value and correlation of $a_{fl}(t)$ are equal to
\begin{eqnarray}\label{field_cond}
    &\braket{a_{fl}(t)} = 0,\\
    &\braket{a_{fl}(t_1)a_{fl}(t_2)} = \frac{(4-\pi)\rho}{m_a^2}\tau\cos^2(m_at_1)\delta(t_1 - t_2),
\end{eqnarray}

The density matrix (\ref{dens}) obeys the following master equation
\begin{equation}\label{start_eqn}
    i\dfrac{d\rho_\nu(t)}{dt} = [H(t), \rho_\nu(t)].
\end{equation}
The presence of classical and fluctuating terms of ALPs field allows to decompose Hamiltonian
\begin{equation}
H(t) = a_{m}(t)J(t) + a_{fl}(t)J(t),
\end{equation}
where 
\begin{equation}
    J(t) = \int d^3 \bfx j(x). 
\end{equation}
The mean field $a_{m}(t)$ can be removed by an additional rotation by operator $U_{m}$ which obeys to the following equation

\begin{equation}
    i\dfrac{d}{dt}U_{m} = a_{m}(t)J(t)U_{m}
    .
\end{equation}
After this rotation $J(t)$ and $\rho_\nu$ get the following form

\begin{eqnarray}
\label{rot_rho}
\rho_\nu' (t) = U_{m}^{\dagger}\rho_\nu(t)U_{m}, \\
\label{rot_h}
J'(t) = U_{m}^{\dagger}J(t)U_{m},
\end{eqnarray}
Eq. (\ref{start_eqn}) also transforms accordingly:
\begin{equation}
    i\dfrac{d\rho_\nu'(t)}{dt} = a_{fl}(t)[J'(t), \rho_\nu'(t)].
\end{equation}
The formal solution of this equation is
\begin{widetext}
\begin{equation}\label{presol}
    \rho_\nu'(t) = \rho_\nu'(t_0) - i\int_{t_0}^t a_{fl}(t_1)[J'(t_1),\rho_\nu'(t_0)] - \int^t_{t_0}dt_1\int^{t_1}_{t_0}dt_2a_{fl}(t_1)a_{fl}(t_2)[J'(t_1),[J'(t_2),\rho_\nu(t_0)]] + ...
\end{equation}
After averaging (\ref{presol}) with (\ref{aver}) it becomes
\begin{equation}\label{presol_aver}
    \rho_\nu'(t) = \rho_\nu'(t_0) - \int^t_{t_0}dt_1\frac{(4-\pi)\rho}{m_a^2}\cos^2(m_at_1)\tau[J'(t_1),[J'(t_1),\rho_\nu(t_0)]] + ...
\end{equation}
\end{widetext}

 For the case of minimal ALPs mass $m_a\approx 10^{-22}$eV the period of ALPs field coherent oscillations is equal to $\frac{2\pi}{m_a} \approx 1.3$ year. The period decreases as axion mass increases.

Due to the fact that the data acquisition period in neutrino experiments usually extend over several years, it is possible to average out the coherent oscillations. So, $\cos^2(m_at) \approx \frac{1}{2}$ and moreover, all the odd terms in (\ref{presol_aver}) vanish. Also, after averaging, $a_m(t)$ becomes zero, and $U_m$ becomes an identity matrix. Therefore, the averaged over data acquisition period  solution \eqref{presol_aver} obeys  

\begin{equation}\label{Redfield}
    \dfrac{d\rho_\nu'(t)}{dt} = -\frac{(4-\pi)\rho}{2m_a^2}\tau[J'(t),[J'(t), \rho_\nu'(t)]].
\end{equation}
Finally, in the Schrodinger representation
\begin{equation}\label{equation_for_rho}
    \dfrac{d\rho_\nu(t)}{dt} = - i[H_0,\rho_\nu(t)] -\frac{(4-\pi)\rho}{2m_a^2}\tau[J,[J, \rho_\nu(t)]],
\end{equation}
where $J = J(0)$, $H_0$ is the Hamiltonian that includes the vacuum Hamiltonian and one that describes neutrino interaction with matter.

The obtained equation (\ref{Redfield}) is the Redfield-type. In the basis of neutrino mass states matrix element of $J$ has the following form
\begin{equation}
    J_{ij} =i( g_V^{ij}\bar u(E_{\bfp i},\vec\bfp)u(E_{\bfp j},\vec\bfp) + g_A^{ij}\bar u(E_{\bfp i},\vec\bfp)\gamma_5u(E_{\bfp j},\vec\bfp)).
\end{equation}
Since the interaction between neutrino and classical ALPs field does not change the neutrino of chirality, we consider the case of the left-chiral neutrinos. Therefore,
\begin{widetext}
\begin{eqnarray}
&\bar u(E_{\bfp i},\vec\bfp)u(E_{\bfp j},\vec\bfp) = \frac{\sqrt{(E_i+m_i)(E_j+m_j)}-\sqrt{(E_i-m_i)(E_j-m_j)}}{2\sqrt{E_iE_j}}, \\
&\bar u(E_{\bfp i},\vec\bfp)\gamma_5 u(E_{\bfp j},\vec\bfp) = \frac{\sqrt{(E_i-m_i)(E_j+m_j)}-\sqrt{(E_i+m_i)(E_j-m_j)}}{2\sqrt{E_iE_j}}.
\end{eqnarray}
\end{widetext}
In the relativistic limit we obtain
\begin{equation}\label{hij}
    J_{ij} \approx i\frac{m_i^2-m_j^2}{2E_\nu F}(C_{V}^{ij}-C_{A}^{ij}).
\end{equation}
As it is seen, in the Hamiltonian of the classical ALPs field the diagonal elements are zero.

The dissipation term 

\begin{equation}
D[\rho_\nu (t)] = -\frac{(4-\pi)\rho}{2m_a^2}\tau[J,[J,\rho_\nu (t)]]
\end{equation}
can be decomposed into SU(3) group generators $\lambda_i$ expressed in terms of Gell-Mann matrices. It can be written as
\begin{equation}
    D[\rho_\nu (t)] = \sum_{a,b =0}^8D_{ab}\rho_a\lambda_b,
\end{equation}
where $\rho_a = 2\Tr(\rho_\nu(t)\lambda_a)$. The matrix element $D_{ab}$ is given by
\begin{equation}
    D_{ab} = 
    - \frac{(4-\pi)\rho}{2m_a^2}\tau \sum _{c,d,e =1}^8 h_c h_d f_{c e b} f_{dea}.
\end{equation}
Here $h_a = 2\Tr(J\lambda_a)$ and $f_{abc}$ is a structure constant of the SU(3) group. For the full dissipation matrix we get
\begin{widetext}
\begin{eqnarray}\label{D}
&D = \nonumber -\frac{(4-\pi)\rho}{8m_a^2}\tau \times\\
&\times \left(
\begin{array}{ccccccccc}
 0 & 0 & 0 & 0 & 0 & 0 & 0 & 0 & 0 \\
0 & 4 h_2^2+h_5^2+h_7^2 & 0 & 0 & 3 h_2 h_5 & 0 & -3 h_2 h_7 & 0 & 2\sqrt{3}h_5h_7 \\
 0 & 0 & h_5^2+h_7^2 & 0 & 0 & -h_2 h_5 & 0 & -h_2 h_7 & 0 \\
 0 & 0 & 0 & 4 h_2^2+h_5^2+h_7^2 & -3 h_2 h_7 & 0 & -3 h_2 h_5 & 0 & \sqrt{3}(h_5^2-h_7^2) \\
 0 & 3 h_2 h_5 & 0 & -3 h_2 h_7 & h_2^2+4 h_5^2+h_7^2 & 0 & 3 h_5 h_7 & 0 & \sqrt{3}h_2h_7 \\
 0 & 0 & -h_2 h_5 & 
 0 & 0 & h_2^2+h_7^2 & 0 & -h_5 h_7 & 0 \\
 0 & -3 h_2 h_7 & 0 & -3 h_2 h_5 & 3 h_5 h_7 & 0 & h_2^2+h_5^2+4 h_7^2 & 0 & -\sqrt{3}h_2h_5 \\
 0 & 0 & -h_2 h_7 & 0 & 0 & -h_5 h_7 & 0 & h_2^2+h_5^2 & 0 \\
 0 & 2\sqrt{3}h_5h_7 & 0 & \sqrt{3}(h_5^2-h_7^2) & \sqrt{3}h_2h_7 & 0 & -\sqrt{3}h_2h_5 & 0 & 3(h_5^2+h_7^2) \\
\end{array}
\right),
\end{eqnarray}
\end{widetext}
where
\begin{eqnarray}
\label{h2}
h_2 = \frac{\Delta m^2_{21} \left(C^{21}_V-C^{21}_A\right)}{E_\nu F}, \\
\label{h5}
h_5 = \frac{\Delta m^2_{31} \left(C^{31}_V-C^{31}_A\right)}{E_\nu F}, \\
\label{h7}
h_7 = \frac{\Delta m^2_{32} \left(C^{32}_V-C^{32}_A\right)}{E_\nu F}.
\end{eqnarray}
Matrix elements $D_{0i}=D_{i0}=0$, which means that there is no neutrino loss under the influence of the ALPs field fluctuations. 

Note, that the obtained dissipative matrix \eqref{D} contains nonzero off-diagonal elements. The obtained particular form of the dissipative matrix follows from the exact account for the type of coupling between neutrino and ALPs, given by (\ref{Hamiltonian})-(\ref{gva}). In experimental searches for neutrino quantum decoherence (see, for example, recent works \cite{KM3NeT:2024jji,Barenboim:2024wdn,Domcke:2025lzg}), only diagonal dissipative matrices are commonly used. As a notable exception, in \cite{Bera:2023ksx,PhysRevD.99.075022, Buoninfante:2020,Capolupo:2018hrp} non-diagonal dissipative matrix was considered and it was shown that the probability of flavour neutrino oscillations depends on the Majorana CP-violation phase if non-diagonal elements of the decoherence matrix are not zero. This may lead to CP violation for Majorana neutrinos. Additionally, in \cite{PhysRevD.99.075022, Buoninfante:2020,Capolupo:2018hrp} it was demonstrated that the neutrino oscillation probability remains invariant under T-transformation, which in sum results in CPT violation for Majorana flavour neutrino oscillations when a non-diagonal dissipative matrix is considered. However, it is worth noting that the obtained dissipative matrix differs from those used in \cite{Bera:2023ksx,PhysRevD.99.075022, Buoninfante:2020,Capolupo:2018hrp}. For instance, in the two-neutrino approximation, the dissipative matrix (\ref{D}) becomes diagonal, unlike those used in the papers mentioned above. 

In the literature (see, for instance, \cite{Giunti:2023,Ternes:2025}), the dependence of the decoherence matrix on the neutrino energy is usually taken in the form 

\begin{equation}\label{DijE}
    D_{ij}(E_\nu) = D_{ij}(E_0)\left(\frac{E_\nu}{E_0}\right)^n,
\end{equation}
where $E_0$ is a pivot energy scale that is usually chosen to be 1 GeV.

In (\ref{DijE}) $n$ is a power-law index that depends on the physics behind. The model with $n=-1$ corresponds to the processes of neutrino decay \cite{MyNew}. The scenario with $n = -2$ gets a natural theoretical justification within the framework of models that take into account the interaction of neutrinos with gravitational waves \cite{Domi:2024}. In the context of quantum gravity effects responsible for the loss of coherence, the positive values of the parameter $n$ are the most expected, although a number of theoretical constructions allow for the possibility of negative values of this parameter. As can be seen from the dependence of the dissipative matrix (\ref{D}) on the neutrino energy, for the case of decoherence of oscillations caused by interaction with fluctuations of axionlike particles, the value of the power index is also $n=-2$.

\section{Conditions of applicability}\label{sec:apply}

Herebelow, we discuss the possibility of using our developed approach for the case of the fluctuating ALPs dark matter and, firstly, we should determine its correlation time. In \cite{Schive:2014dra, Centers:2019dyn, hgnx-w1dn} fluctuations of dark matter were studied (in \cite{Centers:2019dyn} Rayleigh perturbations were considered) and to describe it, authors introduced such a concept as coherence time. In case, when the time of observing dark matter is much shorter than the coherence time, fluctuations are not observed and dark matter oscillates coherently. However, if the observation of dark matter lasts for a much longer period than the coherence time, its density becomes completely random. We assume the possibility of identification of the correlation time with the coherence time. Therefore, the expression for the correlation time becomes the same as for the coherence time \cite{Schive:2014dra, Centers:2019dyn, hgnx-w1dn}
\begin{equation}
    \tau = \frac{1}{m_av^2},
\end{equation}
where $v$ is the ALPs dark matter velocity. As is mentioned above, its order is
$10^{-3}$ c. 

Such fluctuations are described by the finite correlation time, and therefore their correlation should not be described by a delta-function but by (see in \cite{Balantekin})
\begin{eqnarray}
    &\braket{a_{fl}(t_1)a_{fl}(t_2)} = \nonumber \\
    &=\frac{(4-\pi)\rho}{m_a^2}\cos(m_at_1)\cos(m_at_2)\Theta(\tau -|t_1 - t_2|),
\end{eqnarray}
where $\Theta(x)$ is a Heaviside function. For such correlations, the derivation of \eqref{equation_for_rho} should be modified. Consider the second order term of the formal solution \eqref{presol}. After averaging it approximately becomes
\begin{widetext}
\begin{eqnarray}
    &\frac{(4-\pi)\rho}{m_a^2}\int^t_{t_0}dt_1\int^{t_1}_{t_1-\tau}dt_2\cos(m_at_1)[J(t_1),[\cos(m_at_1)J(t_1) + \frac{d(\cos(m_at_1)J(t_1))}{dt}(t_2-t_1),\rho_\nu(t_0)]] = \nonumber \\
    &\frac{(4-\pi)\rho}{m_a^2}\int^t_{t_0}dt_1\int^{t_1}_{t_1-\tau}dt_2[J(t_1),[\cos^2(m_at_1)J(t_1) + \nonumber \\
    & +\left(\cos^2(m_at_1)\frac{dJ(t_1)}{dt}+\cos(m_at_1)\sin(m_at_1)J(t_1)\right)(t_2-t_1),\rho_\nu(t_0)]].
\end{eqnarray}

After averaging over coherent oscillations it becomes


\begin{eqnarray}\label{second_order}
    &\frac{(4-\pi)\rho}{2m_a^2}\int^t_{t_0}dt_1\int^{t_1}_{t_1-\tau}dt_2[J(t_1),[J(t_1) + \frac{dJ(t_1)}{dt}(t_2-t_1),\rho_\nu(t_0)]] = \nonumber \\
    &=\frac{(4-\pi)\rho}{2m_a^2}\left(\tau\int^t_{t_0}dt_1[J(t_1),[J(t_1),\rho_\nu(t_0)]] - \frac{\tau^2}{2}\int^t_{t_0}dt_1[J(t_1),[\frac{dJ(t_1)}{dt},\rho_\nu(t_0)]]\right).
\end{eqnarray}

\end{widetext}

It is possible to rewrite the derivative $\frac{dJ(t_1)}{dt}$ in the following form
\begin{equation}
    \frac{dJ(t_1)}{dt} =  U_0^\dagger[H_0,J]U_0,
\end{equation}
where
\begin{equation}
    U_0 = Texp \left( -i \int^t_{t_0} H_0(t') dt' \right)
\end{equation}
Therefore, in order for the \eqref{second_order} to equal to the averaged secondary order term of \eqref{presol_aver} and, as a consequence, for the solution \eqref{presol} to satisfy \eqref{equation_for_rho}, the following condition must be fulfilled
\begin{equation}
\tau E_{max} \ll 1,
\end{equation}
where $E_{max}$ corresponds to the maximum element of matrix $[H_0,J]$. Therefore

\begin{equation}\label{cond}
    \frac{|\Delta m^2_{13}|^2}{8m_av^2E_\nu^2F} \ll 1.
\end{equation}

When considering axionlike particles as candidates for dark matter, their masses are subject to strict restrictions. Despite the fact that the upper limit of the mass (for the majoron) is 2.8 MeV \cite{PhysRevD.99.096005}, such a value is rarely used and significantly lower values are considered in most theoretical models. In particular, for the wave dark matter scenario, the characteristic mass range lies below 30 eV \cite{annurev}. Since we consider below constrains on decoherence parameter, based on KamLAND data, correlation time shouldn't exceed time of neutrino flight from source to detector, which equals to $\approx $ 0.6 ms \cite{SUZUKI20101}. Therefore, the range of considering ALPs mass lies between $10^{-7}$ eV to 10 eV.

\section{Estimation of neutrino coupling constants with axionlike particles}\label{sec:est}

For numerical estimates, constraints on the decoherence parameter $\gamma_0=D_{ij}(E_0)$ of neutrino oscillations are used, which were obtained in \cite{Giunti:2023} from reactor neutrino experiments data, where the diagonal decoherence matrix without neutrino loss is considered. For the case of $n = -2$, the closest upper limit for our dissipative matrix of $\gamma_0$ is $7.8\times 10^{-27}$ GeV, which was obtained from the analysis of the KamLAND data. Also, based on the assumption that the dark matter density, $\Omega_h$, equals 0.12, the ALPs energy density should be approximately $\rho \approx 0.3 \text{ GeV} / \text{ cm}^{3}$ (see, for instance \cite{Tsutsui:2023jbk}).

We also use the neutrino mixing parameters such as neutrino mass-squared difference $\Delta m^2_{12} = 7.49 \times 10^{-5} \text{ eV}^2$ and $\Delta m^2_{13} = 2.534 \times 10^{-3} \text{ eV}^2$ \cite{Esteban:2024}.

\begin{figure}[h!]
	\centering 
	\includegraphics[scale = 0.6]{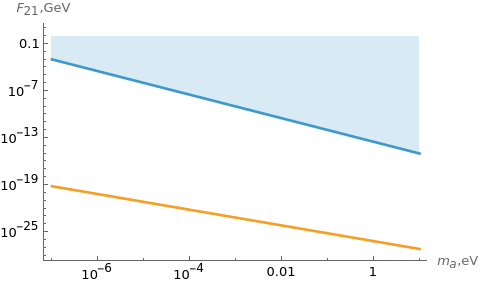}
	\caption{\label{fig:f21} Dependence of the coupling constant of neutrino interaction with axionlike particles $F_{21}$ on the ALPs mass. The blue line corresponds to \eqref{ALPs-neutrino_est}, the orange line to \eqref{F_cond}. The blue area describes allowed range of the coupling constants.}
\end{figure}

\begin{figure}[h!]
	\centering 
	\includegraphics[scale = 0.6]{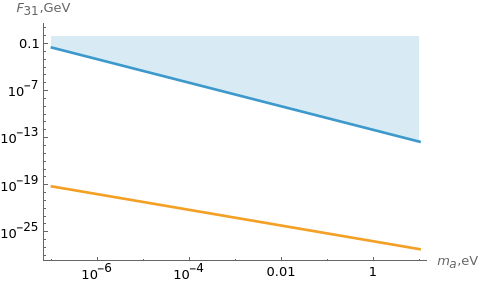}
	\caption{\label{fig:f31} Dependence of the coupling constant of neutrino interaction with axionlike particles $F_{31}$ on the ALPs mass. The blue line corresponds to \eqref{ALPs-neutrino_est}, the orange line to \eqref{F_cond}. The blue area describes allowed range of the coupling constants.}
\end{figure}

\begin{figure}[h!]
	\centering 
	\includegraphics[scale = 0.6]{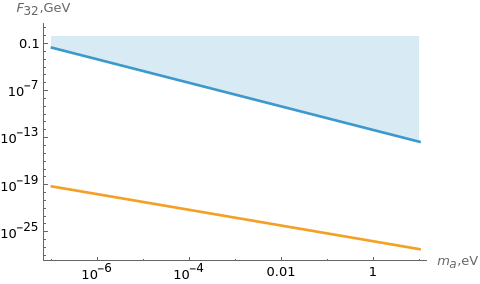}
	\caption{\label{fig:f32} Dependence of the coupling constant of neutrino interaction with axionlike particles $F_{32}$ on the ALPs mass. The blue line corresponds to \eqref{ALPs-neutrino_est}, the orange line to \eqref{F_cond}. The blue area describes allowed range of the coupling constants.}
\end{figure}

To obtain the most conservative estimates of the coupling constants between ALPs and neutrinos, we focused our analysis on the diagonal elements of the dissipative matrix (\ref{D}). In the literature (see, for instance \cite{Giunti:2023, Ternes:2025}) three cases of dissipative matrices are considered:

\begin{eqnarray}
    D_\text{phase-pert.} = \text{diag}(0, \Gamma, \Gamma, 0, \Gamma, \Gamma, \Gamma, \Gamma, 0), \\
    \label{state_select}
    D_\text{state-select} = \text{diag}(0, \Gamma, \Gamma, \Gamma, \Gamma, \Gamma, \Gamma, \Gamma, \Gamma), \\
    D_\text{$\nu$-loss} = \text{diag}(\Gamma, \Gamma, \Gamma, \Gamma, \Gamma, \Gamma, \Gamma, \Gamma, \Gamma),
\end{eqnarray}
where $\Gamma = \gamma_0\left(\frac{E_\nu}{E_0}\right)^n$. The  dissipative matrix (\ref{D}) approximately corresponds to the case \eqref{state_select}, so we can constrain the diagonal elements of the obtained dissipative matrix by the value of $\Gamma$. These conservative estimations lead to the following limits on $h_i^2$:

\begin{eqnarray}
    4h_2^2 + h_5^2 + h_7^2 < \frac{8m_a^2\gamma_0}{(4-\pi)\rho\tau}\left(\frac{E_0}{E_\nu}\right)^2,  \\
    h_2^2 + 4h_5^2 + h_7^2 < \frac{8m_a^2\gamma_0}{(4-\pi)\rho\tau}\left(\frac{E_0}{E_\nu}\right)^2,  \\
    h_2^2 + h_5^2 + 4h_7^2 < \frac{8m_a^2\gamma_0}{(4-\pi)\rho\tau}\left(\frac{E_0}{E_\nu}\right)^2. 
\end{eqnarray}
Then it follows
\begin{eqnarray}
    h_2^2 < \frac{4m_a^2\gamma_0}{3(4-\pi)\rho\tau}\left(\frac{E_0}{E_\nu}\right)^2,  \\
    h_5^2 < \frac{4m_a^2\gamma_0}{3(4-\pi)\rho\tau}\left(\frac{E_0}{E_\nu}\right)^2,  \\
    h_7^2 < \frac{4m_a^2\gamma_0}{3(4-\pi)\rho\tau}\left(\frac{E_0}{E_\nu}\right)^2. 
\end{eqnarray}
From this, we obtain the following bounds on the ALPs-neutrino coupling constants

\begin{equation}\label{ALPs-neutrino_est}
	\frac{F}{C_V^{ij}-C_A^{ij}} > \frac{\Delta m_{ij}^2\sqrt{3(4-\pi)\rho}}{2(m_a)^\frac{3}{2}vE_0\sqrt{\gamma_0}}.
\end{equation}
These conditions are used for derivation of the numerical estimations (see Figs. \ref{fig:f21}-\ref{fig:f32} and Table \ref{neutrino_bounds}).

From the condition of applicability \eqref{cond} of the developed method the coupling constants should obey
\begin{equation}\label{F_cond}
    \frac{F}{C_V^{ij}-C_A^{ij}} > \frac{|\Delta m^2_{13}|^2}{8m_av^2E_\nu^2}.
\end{equation}
Since the limits on the decoherence parameter are obtained from the reactor neutrino experiments, we use the neutrino energy $E_\nu =$  10 MeV. The estimates for the coupling constants of neutrinos with ALPs are illustrated in Figs. \ref{fig:f21}-\ref{fig:f32}. 

Using the ALPs mass $m_a=10^{-7}$ eV, we obtain the lower bounds on the coupling constants $F^{ij}_{V,A}$ of neutrinos with ALPs (see Table \ref{neutrino_bounds}). 

\begin{table}[h]
	\centering
	\begin{tabular}{|c|c|c|}
		\hline
		& V or A & V-A\\
		\hline
		\textrm{$F^{31}_{V,A}$} & \textrm{$3.5\times10^{-2}$}&\textrm{$1.8\times10^{-2}$}\\
		\hline
		\textrm{$F^{32}_{V,A}$} & \textrm{$3.4\times10^{-2}$}&\textrm{$1.7\times10^{-2}$} \\
		\hline
		\textrm{$F^{21}_{V,A}$} & \textrm{$1.0\times10^{-4}$}&\textrm{$0.5\times10^{-4}$} \\
		\hline
	\end{tabular}
	\caption{The lower bounds on the coupling constants of neutrinos with ALPs (in GeV) for the ALPs mass $m_a=10^{-7}$ eV.}
	\label{neutrino_bounds}
\end{table}
It is possible to compare our results with \cite{Cl_ALPs_Neu}, where neutrino oscillations in a classical ALPs field are considered. In that paper, estimates for decay constant $F$ are obtained (they considered coupling constant $g = \frac{1}{F}$) based on the future sensitivities or constraints of the oscillation experiments as well as the astrophysical bounds. The method used in \cite{Cl_ALPs_Neu} allows to study the ALPs mass range between approximately  $10^{-22}$ eV and $10^{-7}$ eV (since the contribution of the ALPs field to the neutrino Hamiltonian is vanishing for $m_a > 10^{-7}$ eV) and gives the value of $F$ between $10^{-2}$ GeV and 1 GeV. Our approach allows one to consider the neutrino-ALPs interaction for the masses of ALPs from $10^{-7}$ eV to $10$ eV. 

Also, effective ALPs-neutrino interaction studied in \cite{Bonilla:2023dtf}, that contributes to the ALP couplings to electroweak gauge bosons at the loop level. In this paper, ALPs mass ranges from $10^5$ eV to $10^{12}$ eV and the sum of the diagonal couplings in a flavor basis is bounded, using the several experiments, such as meson or kaon decays, or the astrophysical data from the supernova explosion SN1987A. Here, the strongest limit on the decay constant $F$ from the experiments on meson and kaon decays is equal to $10^3$ GeV, while the astrophysical bounds give the limit equal to $10^6$ GeV.

Hamiltonian (\ref{Hamiltonian}) also describes the processes of neutrino decay. In a series of works, the constraints on corresponding neutrino decays were obtained \cite{Funcke:2019grs,PhysRevD.72.103514, PhysRevD.84.038701,Berryman:2014qha, SNO:2018pvg, Ascencio-Sosa:2018lbk, Gago:2017zzy, Farzan:2002wx, MyNew}. From here, it is possible to find the bounds on ALP-neutrino coupling scale $F$. However, a direct comparison is not straightforward, as these studies employ distinct physical models.

\section{Conclusion}\label{sec:conc}

We study the neutrino quantum decoherence caused by neutrino interactions with stochastic fluctuations in a classical axionlike particle field. Assuming that the random fluctuations in the ALPs field exhibit a Rayleigh behavior, we demonstrate that the neutrino density matrix for this interaction is governed by the Redfield equation. 

Of particular importance in the study is the explicit calculation of the dissipative matrix (\ref{D}), which made it possible to quantify the influence of ALPs field stochastic fluctuations on the processes of neutrino propagation and oscillations.  An important result is that we have demonstrated the existence of nonzero off-diagonal elements in the dissipative matrix. Usually, in the literature the off-diagonal elements are omitted. 

Using constraints on the decoherence parameters obtained from the analysis of sensitivities of reactor neutrino experiments, as well as considering the region of ALPs masses between $10^{-7}$ eV and $10$ eV of axionlike particles, we have obtained bounds on coupling constants between neutrinos and axionlike particles. These results provide new insights into possible neutrino nonstandard interactions and new physics beyond the Standard Model.

\section*{ACKNOWLEDGMENTS}

This work is supported by the Russian Science Foundation (project No. 24-12-00084).

%
\bibliography{main}
\end{document}